\documentclass[fleqn,twoside]{article}
\usepackage{espcrc2}

\usepackage{graphicx}
\usepackage[figuresright]{rotating}

\newcommand{\lae}{\lower .7ex\hbox{$\;\stackrel{\textstyle <}{\sim}\;$}}
\newcommand{\bc}{\begin{center}}
\newcommand{\ec}{\end{center}}
\newcommand{\beq}{\begin{equation}}
\newcommand{\eeq}{\end{equation}}
\newcommand{\beqa}{\begin{eqnarray}}
\newcommand{\eeqa}{\end{eqnarray}}

\newcommand{\Phat}{\hat{P}}

\newcommand{\AmS}{{\protect\the\textfont2
  A\kern-.1667em\lower.5ex\hbox{M}\kern-.125emS}}

\graphicspath{{./}}

\title{
{\small DESY 04-113 \hfill {\tt hep-ph/0407089}\\
July 2004 \hfill SFB/CPP-04-026}\\
Non--Singlet QCD Analysis of the Structure Function $F_2$ in
3-Loops}

\author{J. Bl\"umlein\address[DSZN]{DESY, Platanenallee 6,
        D--15738 Zeuthen, Germany},
        H. B\"ottcher\addressmark[DSZN]
        and A. Guffanti\addressmark[DSZN]}

\begin{document}

\begin{abstract}
\noindent
First results of a non--singlet QCD analysis of the structure function
$F_2(x,Q^2)$ in 3--loop order based on the non--singlet world data are
presented. Correlated errors are determined and their propagation through
the evolution equations is performed analytically. The value for $\alpha_s(M_Z)$
is determined to be $0.1135 +/- 0.0023/0.0026$ compatible with results from
other QCD analyses. Low moments for $u_v(x)$, $d_v(x)$  and $u_v(x) -
d_v(x)$ with correlated errors are calculated which may be compared with 
results from lattice simulations.
\vspace{1pc}
\end{abstract}

\maketitle

\section{INTRODUCTION}

\vspace*{0.10cm}
\noindent
A consistent 3-loop QCD analysis of the unpolarized structure function
$F_2(x,Q^2)$ can be carried out after having the recently completed
next-to-next-to-leading order (NNLO) anomalous dimensions
available~\cite{MVV} in addition to the 2-loop Wilson
coefficients~\cite{vNV}. Two NNLO QCD analyses have been
performed previously based on structure
function and other hard scattering data~\cite{MRST03}, and
structure function data only~\cite{A02}. In these analyses singlet--
and non--singlet evolution was dealt with in parallel. The present
analysis concentrates
on the non--singlet evolution only in order to firstly obtain an accurate
as possible picture for the valence quark distributions and the value of
the QCD coupling constant $\alpha_s(M_Z^2)$ with correlated errors in
NNLO.
Analyzing only the non--singlet data has the advantage that large gluon
and sea quark effects remain decoupled. A NNLO QCD
analysis of $F_2(x,Q^2)$ allows to reduce the theoretical error in
determining $\alpha_s(Q^2)$ to at least the level of the experimental
error since the factorization and renormalization scale uncertainties
reduce significantly. Comparison of QCD analysis results with results from
recent lattice simulations concerning the low order moments has shown
astonishing agreement in the polarized case~\cite{BB02}. With the
steadily improving lattice calculations this might also become
feasible in the unpolarized case, where systemantic effects have still 
been large during the last years.  In this letter we describe the results
of an analysis of the deeply inelastic non--singlet world data for charged
lepton--nucleon scattering.

\section{QCD FORMALISM}

\vspace*{0.10cm}
\noindent
In {Mellin--$N$} space the non--singlet (NS) parts of a structure
function $F_i(N,Q^2)$ are given by 
\beqa
F_i^{\pm,{\rm v}}(N,Q^2) & = & \left[1+C_{1i}(N)a+C_{2i}(N)a^2\right] 
\nonumber \\ 
                         &   & f^{\pm,{\rm v}}(N,Q^2)\,,  
\eeqa

\noindent
where the $C_{ki}(N)$ are the corresponding Wilson coefficients and
$f^{\pm,{\rm V}}(N,Q^2)$ are the non--singlet quark combinations. The 
symbol $a$ 
denotes the strong coupling constant  normalized to $a(Q^2) =
\alpha_s(Q^2) / 4\pi$. The quark combinations to be considered are 

\beq
\Delta^{\pm} = (u \pm \bar{u}) - (d \pm \bar{d})~,
\eeq

\beq
v = (u - \bar{u}) + (d - \bar{d})~.
\eeq

\noindent
The non--singlet parts of $F_2$ are proportional to 

\beqa
F_2^{NS} \propto \frac{1}{3} \Delta^+, & F_{2}^{p,v} \propto
\frac{5}{18}v + \frac{1}{6}\Delta^-,\quad {\rm and} & \nonumber \\
F_{2}^{d,v} \propto \frac{5}{18}v\,. & &
\eeqa   

\noindent
$F_2^{NS}$ stands for the difference of proton and deuteron 
data in the
range $x < 0.3$ while the other combinations are used in the valence
approximation for $x > 0.3$. All these combinations evolve as $+$-combinations
in $Q^2$. 
The
relevant parton densities to be determined in a non--singlet QCD analysis
are $xu_v(x,Q^2)$, $xd_v(x,Q^2)$, and $x(\bar{d} - \bar{u})(x,Q^2)$.

The solution of the evolution equation to 3-loops reads for the valence
distribution as a example
\beqa
  V(Q^2) & = & V(Q_0^2) 
       \left(\frac{a}{a_0}\right)^{-\Phat_0/\beta_0} \nonumber \\ 
   &&\Biggl\{1-\frac{1}{\beta_0}(a-a_0)
     \Biggl[\Phat^\pm_1-\frac{\beta_1}{\beta_0}\Phat_0\Biggr]\Biggr. 
       \nonumber \\     
   &&\Biggl.-\frac{1}{2\beta_0}\left(a^2-a_0^2\right)\Biggr.
     \Biggl[\Phat_2^{\pm,{\rm V}}-\frac{\beta_1}{\beta_0}\Phat^\pm_1
     \Biggr.   \nonumber\\
   &&\Biggl.+\Biggl(\frac{\beta_1^2}{\beta_0^2}-\frac{\beta_2}{\beta_0}
     \Biggr)\Phat_0\Biggr]
       \\ 
   &&+\frac{1}{2\beta_0^2}(a-a_0)^2 
   \Biggl(\Phat^\pm_1-\frac{\beta_1}{\beta_0}\Phat_0\Biggr)^2 \Biggr\} \,,
\nonumber
\eeqa
\noindent
where the $\Phat_i$ are the  splitting 
functions in Mellin space.

\section{PARAMETERIZATION}

\vspace*{0.10cm}
\noindent
The parameterizations of the above mentioned parton densities at the
input scale of $Q^2_0 = 4.0~{\rm GeV^2}$ are determined as

\beqa
x u_v(x,Q_0^2) & = & A_{u_v} x^{a_u} (1 - x)^{b_u} \nonumber \\ 
        & & (1 - 1.108x^{\frac{1}{2}} + 26.283x)\,, 
\eeqa
and

\vspace*{-0.25cm}
\beqa
x d_v(x,Q_0^2) & = & A_{d_v} x^{a_d} (1 - x)^{b_d} \nonumber \\
        & & (1 + 0.895x^{\frac{1}{2}} + 18.179x)\,.
\eeqa

\noindent
Here the values for the coefficients in the polynomial given as numbers
in the intermediary range of $x$  are obtained by a fit and are then kept
fixed since their respective errors are still large. Furthermore, the
analysis requires different $\overline{u}$ and $\overline{d}$ quark
densities. We adopted the choice \cite{MRST02} which gives a good
description of the $(\bar{d}-\bar{u})$ data from E866~\cite{E866}

\beqa
x (\bar{d}-\bar{u})(x,Q_0^2) & = & 1.195 x^{1.24} (1-x)^{9.10}
\nonumber \\ 
        & & (1 + 14.05x - 45.52x^2)\,.
\eeqa

\noindent
The normalization constants $A_{u_v}$ and $A_{d_v}$ are fixed by
the conservation of the number of valence quarks: $\int_0^1 u_v(x) dx
= 2\,,\,\int_0^1 d_v(x) dx = 1$.  

The remaining four  parameters are  determined in the fit for $u_v$ and
$d_v$, the low--$x$ and high--$x$
parameters $a$ and $b$, respectively, along with
$\Lambda_{QCD}$.

\begin{figure}[t]
\includegraphics*[scale=0.40]{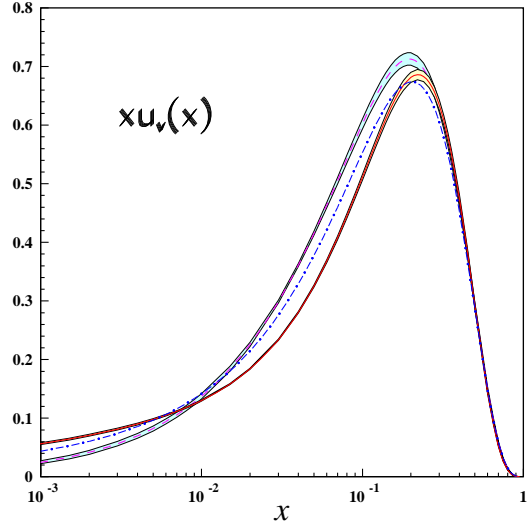}
\vspace*{-0.80cm}
\caption{The parton density $xu_v$ at the input scale
$Q_0^2 = 4.0~{\rm GeV^2}$ (solid line) compared to results obtained by
MRST   (dashed--dotted line)~\cite{MRST03} and  Alekhin (dashed
line)~\cite{A02}. The shaded areas
represent the fully correlated $1\sigma$ statistical error bands.}
\label{fig:pdf}
\end{figure}
\begin{figure}[t]
\includegraphics*[scale=0.40]{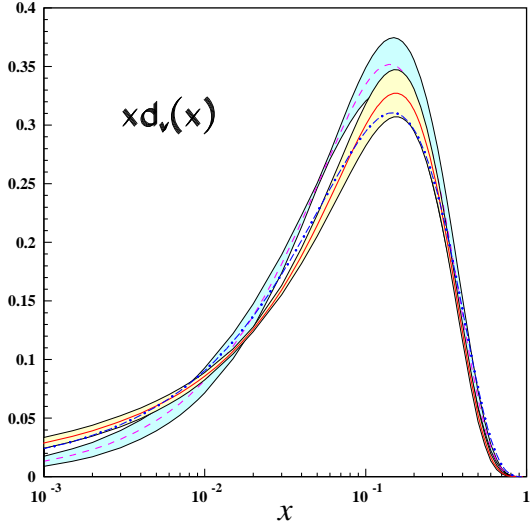}
\vspace*{-0.80cm}
\caption{~The parton density $xd_v$ at the input scale
$Q_0^2$ with the same conditions as in Figure 1a.}
\label{fig:pdf2}
\end{figure}

\section{DATA}

\vspace*{0.10cm}
\noindent
The results presented here are based on 762 data points for the
structure function $F_2(x,Q^2)$ measured on proton and deuteron targets.
The experiments contributing to the statistics were~: BCDMS~\cite{BCDMS},
SLAC~\cite{SLAC}, NCM~\cite{NMC}, H1~\cite{H1}, and
ZEUS~\cite{ZEUS}. The BCDMS data were recalculated replacing $R_{QCD}$
by $R_{1998}$~\cite{R1998}. All deuteron data were corrected for Fermi
motion and offshell effects~\cite{MT}. The non--singlet structure
function $F_2^{NS}$ was constructed according to the relation $F_2^{NS} =
2 (F_2^p - F_2^d)$ from proton and deuteron data given at the same $x$
and $Q^2$ values. The region $x > 0.3$ which was selected for
$F_2^{p,d}$ is expected to be dominated by valence quark contributions
while for the region $x < 0.3$ $F_2^{NS}$ data were used. In addition,
a $Q^2$ region of $4.0 < Q^2 < 30000~{\rm GeV^2}$ was chosen and a cut in
the hadronic mass $W^2 > 12.5~{\rm GeV^2}$ was applied in order to
diminish higher twist effects.

In the fitting procedure we allowed for a relative normalization shift
between the data sets within the overall normalization uncertainties
quoted by the experiments or assumed accordingly. These normalization
shifts were fitted once and then kept fixed.

\section{CORRELATED ERROR CALCULATION}

\vspace*{0.10cm}
\noindent
The systematic errors which are known to be partly correlated are not
treated separately in this analysis. For all data sets we used the
simplest procedure by adding the statistical and the total systematic
errors in quadrature being aware of the fact that this will eventually
overestimate the errors.

The fully correlated 1$\sigma$ experimental error bands are determined
via Gaussian error propagation demanding that only fits with a
positive definite covariance matrix are accepted. The gradients of the
parton distributions with respect to the variable parameters needed
here can be calculated analytically at the input scale $Q_0^2$. Their
values at $Q^2$ are given by evolution in { Mellin--$N$} space.

\vspace*{-3mm}
\section{RESULTS}

\noindent
The fit results are summarized in table 1.  

\vspace*{+0.10cm}
\noindent
\renewcommand{\arraystretch}{1.3}
\bc
\begin{tabular}{|c|c|c|}
\hline
$u_v$      & $a$         &  0.314 $\pm$ 0.007 \\
           & $b$         &  4.199 $\pm$ 0.032 \\
\hline
$d_v$      & $a$         &  0.413 $\pm$ 0.047 \\
           & $b$         &  6.196 $\pm$ 0.332 \\
\hline
\multicolumn{2}{|c|}{$\Lambda_{QCD}^{(4)}$} & 227 $\pm$ 30 $MeV$ \\
\hline \hline
\multicolumn{3}{|c|}{$\chi^2 / ndf$ = 652/757 = 0.86} \\
\hline
\end{tabular}
\ec
\renewcommand{\arraystretch}{1.0}

\vspace*{+0.10cm}\noindent
{\sf Table~1: Parameter values of the NNLO QCD fit based on the 
non--singlet world data on $F_2^{em}(x,Q^2)$.} 
\vspace*{+0.25cm}

The value of $\Lambda_{QCD}^{(4)}$ is quite stable against a variation
of the $Q^2$ cut on the data when varying it between $4.0~{\rm GeV^2}$
to $10.0~{\rm GeV^2}$. From the fitted value of $\Lambda_{QCD}^{(4)}$
the following value of $\alpha_s$ is extracted: \\

\bc
$\alpha_s(M_Z^2) = 0.1135 \begin{array}{c} +0.0023 \\ -0.0026 \end{array}$
(expt).
\ec

\noindent
This value is within the errors in good agreement with results from
other NLO/NNLO QCD analyses, see Table~2,
and with the latest value for the world average of
$0.1182 \pm 0.0027$~\cite{Bethke}.

\vspace{3mm}\noindent
The resulting parton densities $xu_v(x)$ and $xd_v(x)$ at the input
scale of $Q_0^2 = 4.0~{\rm GeV^2}$ are presented in
Figures 1, 2. Comparison with results from global analyses
shows
satisfactory agreement. While $xu_v$ is rather well determined, the
error band for $xd_v$ is a bit broader as also found by the other
analyses. 

\vspace{4mm}\noindent
{\footnotesize
\begin{tabular}{|l|lll|c|}
\hline
                  & $\alpha_s(M_Z^2)$ & expt & theory & Ref. \\
\hline
\hline
 NLO   & & & & \\
\hline
 CTEQ6  & 0.1165 & $\pm$0.0065 &                          & [17] \\
 MRST03 & 0.1165 & $\pm$0.0020 & $\pm$0.0030              & [3] \\
 A02    & 0.1171 & $\pm$0.0015 & $\pm$0.0033              & [4] \\
 ZEUS   & 0.1166 & $\pm$0.0049 &               & [19] \\
 H1     & 0.1150 & $\pm$0.0017 & $\pm$0.0050  & [12]
\\
 BCDMS  & 0.110  & $\pm$0.006 &                          & [9] \\
\hline
 BB (pol)     & 0.113 & $\pm$0.004 & \hspace*{-0.50cm}
\footnotesize{$\begin{array}{c} +0.009 \\ -0.006 \end{array}$}  & [5]
\\
\hline
\hline
 NNLO  & & & & \\
\hline     
 MRST03       & 0.1153 & $\pm$0.0020 & $\pm$0.0030  & [3] \\
 A02          & 0.1143 & $\pm$0.0014 & $\pm$0.0009  & [4] \\
 SY01(ep)     & 0.1166 & $\pm$0.0013 &              & [19] \\
 SY01($\nu$N) & 0.1153 & $\pm$0.0063 &              & [19] \\
\hline
\end{tabular}
}

\vspace{3mm}\noindent
{\sf Table~2: Comparison of NLO and NNLO results on $\alpha_s(M_Z^2)$
from deep inelastic scattering, including also global fits.
The NLO H1 value is subject to an additional error of
$+0.0009/-0.0005$
and the NLO ZEUS value of $\pm 0.0018$ due to model dependence.
\normalsize}

\vspace*{3mm}

\noindent
Low  moments for $u_v,~d_v$ and $u_v-d_v$ with correlated
1$\sigma$ errors have been calculated.
An interesting comparison can be made for the second moment of 
$u_v - d_v$ with a result from recent lattice simulations. A value of
$0.180 \pm 0.005$ was derived here and can be compared with 
results of upcoming  lattice simulations, cf. e.g. \cite{GSch}. 
This comparison should be performed for higher moments in the future
as well as for the moments of the individual distributions  $u_v$ and 
$d_v$.

\section{CONCLUSIONS}

\vspace*{.1cm}
\noindent
A non--singlet QCD analysis of the structure function $F_2(x,Q^2)$ based
on the non--singlet world data has been performed at 3--loop order. The
value
determined for $\alpha_s(M_Z^2)$ is compatible within the errors with
results from other QCD analyses and with the world average. New
parameterizations of the parton densities $u_v$ and $d_v$ including
their fully correlated 1$\sigma$ errors are derived. Low order
moments for $u_v,~d_v$ and $u_v-d_v$ are calculated from this analysis.
It will be interesting to compare them with upcoming results from lattice
simulations.
 
\vspace{2mm}\noindent
{\bf Acknowledgment.} This paper was supported in part by DFG 
Sonderforschungsbereich Transregio 9, Computergest\"utzte Theoretische 
Physik and EU grant HPRN--CT--2000--00149.


\begin{thebibliography}{9}
%
\bibitem{MVV}
S.~Moch, J.~A.~M.~Vermaseren, and A.~Vogt, Nucl. Phys.
{\bf B688} (2004) 101. 
%
\bibitem{vNV}
W.~L.~van~Neerven and A.~Vogt, Nucl. Phys. {\bf B568} (2000) 263 and
references therein.
%
\bibitem{MRST03}
A.~D.~Martin et al., hep-ph/0307262.
%
\bibitem{A02}
S.~Alekhin, Phys. Rev. {\bf D68} (2003) 014002. 
%
\bibitem{BB02}
J.~Bl\"umlein and H.~B\"ottcher, Nucl. Phys. {\bf B636} (2002) 225.
%
\bibitem{MRST02}
A.~D.~Martin et al., Eur. Phys. J.{\bf C23} (2002) 73.
%
\bibitem{E866}
R.~S.~Towell et al., Phys. Rev. {\bf D64} (2001)
052002. 
%
\bibitem{BCDMS}
A.~C.~Benvenuti et al., Phys. Lett. {\bf B237} (1990) 592;
Phys. Lett. {\bf 223} (1989) 485. 
%
\bibitem{SLAC}
L.~W.~Whitlow et al., Phys. Lett. {\bf B282} (1992) 475.
%
\bibitem{NMC}
M.~Arneodo et al., Nucl. Phys. {\bf B483} (1997) 3.
%
\bibitem{H1}
C.~Adloff et al., Eur. Phys. {\bf C21} (2001) 33;
Eur. Phys. {\bf C30} (2003) 1. 
%
\bibitem{ZEUS}
J.~Breitweg et al., F. Phys. {\bf C7} (1999) 609;
S.~Chekanov et al., Eur. Phys. {\bf C21} (2001) 443.
%
\bibitem{R1998}
K.~Abe et al., Phys. Lett. {\bf B452} (1999) 194.
%
\bibitem{MT}
W.~Melnitchouk and A.~W.~Thomas, Phys. Lett. {\bf B377} (1996) 11;
Phys. Rev. {\bf C52} (1995) 3373.
%
\bibitem{Bethke}
S.~Bethke, these  Proceedings; J. Phys. {\bf G26} (2000) R27.
%
\bibitem{GSch}
G.~Schierholz, these Proceedings and private communication.
%
\bibitem{CTEQ}
J. Pumplin et al., JHEP 0207:012 (2002).
\bibitem{ZPRD}
S. Chekanov et al., Phys. Rev. {\bf D67} (2003) 012007.
\bibitem{SY}
J. Santiago and F.J. Yndurain, Nucl. Phys. {\bf B611} (2001) 447.
\end{thebibliography}
\end{document}